\documentclass[fleqn,usenatbib]{mnras}

\usepackage{newtxtext,newtxmath}
\usepackage[T1]{fontenc}

\DeclareRobustCommand{\VAN}[3]{#2}
\let\VANthebibliography\thebibliography
\def\thebibliography{\DeclareRobustCommand{\VAN}[3]{##3}\VANthebibliography}

\usepackage{threeparttable}
\usepackage{subcaption}
\usepackage{soul}
\usepackage{graphicx}	% Including figure files
\usepackage{amsmath}	% Advanced maths commands

\title[Radio continuum emission from a tidal dwarf galaxy]{Radio continuum emission from a tidal dwarf galaxy }

\author[B. M. Moncada-Cuadri et al.]{
Blanca M. Moncada-Cuadri,$^{1}$\thanks{E-mail: bmc56@bath.ac.uk}
Ute Lisenfeld,$^{2,3}$
Miguel Querejeta$^{4}$
and Carole Mundell$^{1,5}$
\\
$^{1}$Department of Physics, University of Bath, Claverton Down, Bath, BA2 7AY, UK\\
$^{2}$Departamento de F\'isica Te\'orica y del Cosmos, Universidad de Granada, 18071 Granada, Spain\\
$^{3}$Instituto Carlos I de F\'isica Te\'orica y Computacional, Facultad de Ciencias, 18071 Granada, Spain\\
$^{4}$Observatorio Astronómico Nacional, Alfonso XII, 3, 28014 Madrid, Spain\\
$^{5}$European Space Astronomy Centre (ESAC), Madrid, Spain
}

\date{Accepted 2024 May 31. Received 2024 May 31; in original form 2023 December 25}

\pubyear{2024}

\begin{document}
\label{firstpage}
\pagerange{\pageref{firstpage}--\pageref{lastpage}}
\maketitle

\begin{abstract}
Tidal Dwarf Galaxies (TDGs) form in the debris of galaxy mergers, making them ideal testbeds for investigating star formation in an extreme environment. We present radio continuum EVLA observations spanning 1-2 GHz of the interacting system Arp 94, which contains the TDG J1023+1952. We detect extended radio continuum emission from the disk of the TDG’s putative parent galaxy, the spiral NGC~3227. The TDG lies in front of the spiral disk, partially overlapping in projection. This challenging alignment complicates the separation of the respective contributions of radio emission from the TDG and disk. However, we show that the radio continuum appears more prominent around the TDG's location, suggesting the detection of emission from the TDG. Quantifying this argument, we derive an upper limit of 2.2 mJy for the whole TDG's emission. Our derived inband spectral index map of the system generally shows the expected behaviour of combined thermal and synchrotron radio emission in a galaxy disk, except for a region at the periphery of the disk and the TDG with a flat spectrum (spectral index $\sim$-0.4) unrelated to regions with high H$\alpha$ emission. We speculate that at this location - which coincides with the intersection of faint tidal tails - the collision of gas clouds produces shocks which reaccelerate cosmic ray electrons, and thereby enhance the radio emission. Overall, this study provides new insights about the Arp 94 system and expands the sample of TDGs studied at radio frequencies, with only two confirmed detections so far. 
\end{abstract}


%%%%%%%%%%%%%%%%%%%%%%%%%%%%%%%%%%%%%%%%%%%%%%%%%%

%%%%%%%%%%%%%%%%% BODY OF PAPER %%%%%%%%%%%%%%%%%%

\section{Introduction}

When galaxies interact, tidal forces redistribute galactic material and produce long tidal tails composed of stars and gas. \citet{Zwicky} suggested that a new type of object can form after the gravitational collapse of the tidal debris: tidal dwarf galaxies (TDGs). TDGs exhibit distinct characteristics compared to "classical dwarfs", making them particularly well-suited for exploring molecular gas content and the laws governing star formation.

TDGs are recycled objects, inheriting the metallicity of the discs of their parent galaxies, which means that they possess near-solar metallicity. This metallicity characteristic allows for the reliable use of CO as a tracer for molecular gas emission. Additionally, TDGs are expected to possess minimal amounts of dark matter due to tidal forces predominantly expelling material from their parent galaxies' disks \citep{Duc2012,Lelli2015}. TDGs have often high gas fractions and their strong tidal forces can enhance star formation \citep{Braine2001}. The reduced dark matter content of TDGs, along with their small number of preexisting stars imported from the parent galaxy disc, provides a more direct laboratory for testing theories of star formation in an extreme dynamical environment. In addition, the rates at which TDGs are formed and can survive are still poorly constrained, despite their significant cosmological implications \citep{Ploeckinger2015,Ploeckinger2018}. 

J1023+1952 is a gas-rich TDG located in the interacting system Arp 94. This system is optically dominated by NGC 3227, a SAB(s) pec barred spiral Seyfert galaxy \citep{Mundell1995a}, and NGC 3226, its E2 pec elliptical companion \citep{Rubin1968}. We assume a distance to the system of 14.5 $\pm$ 0.6 Mpc \citep{Yoshii2014}. Studies of the atomic and molecular gas in the system (\citealt{Mundell2004,Querejeta2021}) indicate that the TDG J1023+1952 is kinematically distinct from NGC 3227. Positioned within the northern tidal tail, the TDG partially overlaps the disc of NGC 3227 in projection. Notably, the TDG appears to be located in front of the spiral galaxy, as its presence obscures the stellar light originating from NGC 3227 (\citealt{Mundell1995}, \citeyear{Mundell2004}). 

\begin{figure}
   \centering
 \includegraphics[width=\hsize]{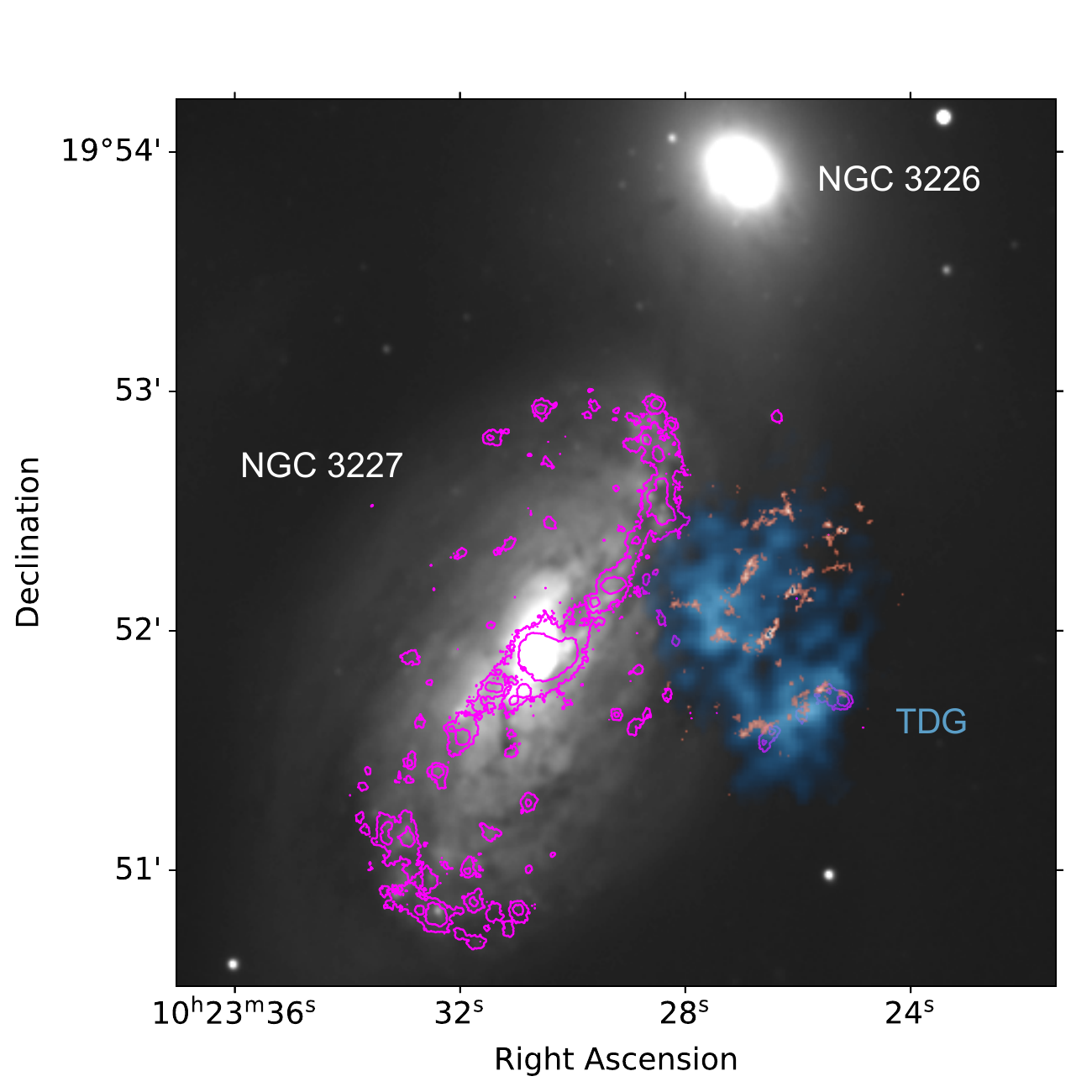}
      \caption{False-colour image of the Arp 94 system combining a $B$-band image (grey scale), and HI intensity (blue) from the VLA \citep{Mundell2004}. We overlay CO(2-1) high-resolution observations from ALMA (orange, \citealt{Querejeta2021}) and H$\alpha$ intensity contours (magenta, \citealt{Mundell2004})
              }
         \label{FigIntro}
         
   \end{figure}

Two facts firmly support the classification of J1023+1952 as a TDG, rather than a pre-existing dwarf galaxy within the system: 1) \citet{Mundell2004} identified a cluster of blue star-forming knots visible through H$\alpha$ emission (see Fig. \ref{FigIntro}), located in the southern region of the cloud. These knots exhibit luminosities ranging from $10^6$ to $5 \times 10^6\  L_\odot$ and display velocities closely matched with the HI cloud, confirming their dissociation from the parent galaxy \citep{Mundell2004}. 2) The star-forming knots exhibit near-solar metallicity (12 + log(O/H) = 8.6, \citealt{Lisenfeld2008}),  significantly higher than what would typically be expected for a dwarf galaxy of similar luminosity \citep{Richer1995}. This metallicity level closely resembles that of NGC 3227, further supporting its classification as a TDG. However, despite these arguments, firmly classifying it as a TDG remains challenging, primarily due to a lack of observational evidence demonstrating that it is a gravitationally bound object (a difficulty that it shares with most TDG candidates). Nonetheless, this uncertainty does not affect the relevance of the scientific questions addressed in this study.

The TDG J1023+1952 exhibits a notable abundance of atomic and molecular gas ($M_{\text{HI}}=8.4\times10^7 M_\odot$, $M_{\text{H}_2}=8.6\times10^7 M_\odot$, \citealt{Mundell2004,Querejeta2021}). While the surface densities of atomic and molecular gas are relatively similar across the extensive 8.9 $\times$ 5.9 kpc \citep{Mundell2004} area covered by the TDG, the formation of stars is predominantly happening in the southern half, along a 2 kpc long ridge (see Fig. \ref{FigIntro}). There is a possibility that star formation is also happening in the north, but it would be extremely limited compared to the south. Consequently, it is possible to identify two different environments within the TDG: a quiescent north and a star-forming south, separated at Dec $\approx$ 19\textdegree 51' 50".

Furthermore, \citet{Querejeta2021} combined high-resolution observations from the 12m array with 7m + TP ALMA observations, thereby recovering emission across all scales. This combined analysis unveiled a remarkably high fraction of diffuse CO emission originating from scales spanning several kiloparsecs, ranging between 80\% and 90\% of the total molecular gas in the TDG. This fraction is considerably higher than the values found in other nearby galaxies \citep{Pety2013,caldu_primo2015}. Therefore, only 10-20\% of the molecular gas is locked up in GMCs, whereas the majority is in a diffuse molecular gas phase. 

TDGs, and J1023+1952 in particular, present a fantastic opportunity to study the interplay between environmental properties and the mechanisms governing the initiation and regulation of star formation beyond our galaxy. Given that star-forming galaxies emit thermal (free-free) and non-thermal (synchrotron) radiation at radio frequencies, the analysis of radio continuum emission allows us to probe various star formation mechanisms and to trace the evolution of star formation throughout the system, from regions where young star-forming blue knots have been identified, to synchrotron emitting cosmic-ray electrons (CRes) accelerated in supernova remnants. Nonetheless, CRes can also be accelerated in interstellar shocks, as seen in Stephan’s Quintet \citep{Xu2003}, or colliding galaxies, such as the Taffy galaxies \citep{Condon1993,Lisenfeld2010}.

 The study of radio continuum emission within TDGs remains a relatively new field of study. To date, radio continuum emission has only been successfully detected in SQ-A and SQ-B, two TDGs located in Stephan’s Quintet galaxy group \citep{Xu2003}. The radio emission detected in SQ-A has a non-thermal character, as evidenced by its spectral index, $\alpha = 0.8 \pm 0.3$. In the case of SQ-B, further investigations by \citet{wroczy2013} unveiled partially polarized radio emission, indicating the presence of a relatively strong magnetic field, $\approx 6.5 \mu G$. This value of the magnetic field is similar to those found in spiral galaxies (see \citealt{Beck2015}), with a significant ordered component ($3.5 \pm 1.2 \mu$G). We must mention that with the exception of starburst dwarf galaxies, dwarf galaxies typically exhibit weaker magnetic fields than those observed in spiral galaxies, often registering at just a few microgauss \citep[e.g.][]{Chyzy2011}.

Additionally, a study by \citet{wrocy2019} hints at the presence of radio emission in TDG HCG9li, part of the Hickson Compact Groups. Should this emission indeed originate within the TDG, it would suggest the existence of a magnetic field as strong as $11-16 \mu G$. However, we must acknowledge the disparities between the radio-derived star formation rate and the rate determined using H$\alpha$ data, where the former exceeds the latter by two orders of magnitude, shedding doubts on the reality of the detection. What remains clear is that the dataset of TDGs with radio continuum detections remains small, hindering our ability to draw definitive conclusions regarding the properties, role of magnetic fields, and the nature of radio emission within TDGs. Therefore, additional observations within TDGs are imperative to achieve a more comprehensive understanding of the star formation process in these unique galaxies.

In this paper, we present the results of investigating the radio continuum emission in the Arp 94 system, where the TDG J1023+1952 resides. We analyse archival data from the \emph{Karl G. Jansky} Expanded Very Large Array (EVLA) between 1 - 2 GHz, broadening the sample of TDGs studied at radio wavelengths. The paper is structured as follows: Section \ref{data} provides an overview of the EVLA dataset (Sect. \ref{observations}), including the data processing (Sect. \ref{calibration} and \ref{imaging}), and the ancillary data (Sect. \ref{ancillary}). In Section \ref{results} we present the radio maps obtained. Section \ref{discussion} discusses different arguments in favour and against the possible detection of radio emission in the TDG. Finally, Section \ref{conclusions} summarises our paper.

\section{Observations and Data Reduction}\label{data}

    \subsection{Observations}\label{observations}

Observations of the Arp 94 system were obtained with the NRAO EVLA under project ID: 13A-066. These observations were carried out in the L-band frequency range (1-2 GHz) using the C-configuration during two separate observing runs in July and August of 2013. 

The L-band is divided into 16 spectral windows (spw) of 64 MHz width, which cover 64 channels of 1 MHz width. A summary of the observations is provided in Table \ref{tab:vla}. 

The observing strategy contained the following steps: scan of an NRAO primary calibrator to calibrate the flux scale, bandpass solutions and polarization angle. This is followed by a scan of a primary low-polarization leakage calibrator; and finally,  alternating scans of the target source, TDG J1023+1952, and the complex gain calibrator. The observation runs were concluded with additional scans of the primary flux calibrator and the polarization leakage calibrator. For further details of the calibrators used see Table \ref{tab:vla}.

\begin{table*}
 \begin{threeparttable}
\centering
\caption{\label{tab:vla} Summary of VLA Observations}
\begin{tabular}{lcccccc}
\noalign{\smallskip} \hline \noalign{\medskip}
Project code &  Band  &  $t_{\text{OS}}$\tnote{a}   & Flux Cal. & Gain Cal. & Working Resolution\\ & & (min) & Name & Name & (arcsec) \\
\noalign{\smallskip} \hline \noalign{\medskip} 
13A-066 &  L & 133  &  3C286 & J1021+2159 &  20 $\times$ 15\\
\noalign{\smallskip} \hline \noalign{\medskip}
\end{tabular}
\begin{tablenotes}
\item [a] Total time on source, adding July and August observing runs
\end{tablenotes}
  \end{threeparttable}
\end{table*}     \subsection{Calibration}\label{calibration}

To calibrate the radio data, we used the Common Astronomy Software Applications (CASA\footnote{http://casa.nrao.edu/}) package (version 6.5.3.28; \citealt{TheCASATeam_2022}). For both datasets, we run the VLA Calibration Pipeline (version 6.2.1.7) as a starting point for the calibration. Subsequently, we conducted a visual inspection of the data to identify and remove any radio frequency interference (RFI) that may not have been flagged by the pipeline. Additionally, we flagged the first and last 5 channels of each spw. Following this, we manually calibrated each dataset following standard procedures in CASA. We started performing the delay and bandpass calibration. Next, we derived corrections for the complex amplitude and phase antenna gains. Finally, we scaled the amplitude gains using our flux density calibrator. We concluded the manual calibration with two rounds of phase self-calibration. 

It is important to note that the dataset obtained in August exhibited lower quality compared to the July observations due to atmospheric and antenna issues. The dataset needed additional flagging at short baselines, which affected its sensitivity to extended emission. In addition, we observed a lack of sufficient uv coverage for the target and the gain calibrator in 3 of the 16 spws of the dataset. Consequently, we were unable to obtain gain solutions for these affected data.

Finally, to enhance uv coverage and improve the signal-to-noise ratio, we combined the July and August observations into a single measurement set. However, it is worth noting that we did not combine the July data with the three spws from the August dataset where gain solutions could not be derived. This precaution was taken to prevent potential contamination of the dataset with data that had not undergone proper calibration procedures.

     \subsection{Imaging}\label{imaging}

We generated the final radio continuum images of TDG J1023+1952 using the CLEAN deconvolution algorithm within the CASA software. For this task, we employed the Multi-Scale, Multi-Term, Multi-Frequency Synthesis (MS-MT-MFS) algorithm \citep{Rau2011} which accounts for the different scales that constitute the sky brightness distribution, as well as the changes of the primary and synthesized beams with frequency. We selected cleaning scales equal to 0 (representing point sources), 1 and 3 times the width of the synthesized beam. 

During the imaging process, we found a significant number of bright sources situated on the edges of the images, which created artifacts in our target source. To reduce the impact of these artifacts, we used the ``pblimit'' parameter within the CLEAN task to produce images much larger than the primary beam so that we could include pixels located on the edges of the field of view. Additionally, we created a mask encompassing all the bright sources within the images, which we used during the cleaning process. 

To later derive the spectral index map across the system (as detailed in Section \ref{spectral index}) we individually processed each spw of the dataset. To ensure a uniform resolution across all maps, we applied different tapers and weights to obtain a final synthesized beam that is similar for each spw. We opted for an intermediate resolution of 20'' x 15'' ($\approx$ 1.2 kpc at our assumed distance) as the common synthesized beam size for all spws, which avoids applying heavy tapers and uniform weights to the spws with higher and lower frequencies respectively. We then selected the weights and tapers for each spw so they would produce images with a resolution close to the selected common one of 20'' x 15''. The specific taper and weight configurations employed for each spw are provided in Table \ref{tab:weights}. Furthermore, we applied elliptical Gaussian smoothing to ensure that all images precisely match the 20'' x 15'' joint resolution. 

After completing the imaging process, we employed the PBCORR task within CASA to correct the final maps for the primary beam effects. The rms noise values for the resultant maps are included in Table \ref{tab:weights}. 
We used the Python algorithm BANE, from the AegeanTools module \citep{Hancock2012} to estimate the spatially changing background noise. This algorithm operates by computing the mean and standard deviation across a grid of pixels, defining a boxed region, and subsequently interpolating these values to generate the final noise and background images. To prevent contamination from source pixels, it implements a 3-sigma clipping process. The unmasked pixels within the designated region are then used to compute both the standard deviation and the median, which serve as our estimates for the noise and the background, respectively. We have used the default grid and box size settings for BANE, as they produced satisfactory results for our images. The estimation of background noise enables us to generate signal-to-noise ratio maps that account for local variations in the noise within the images, which may arise from residual artifacts.

\begin{table}
\begin{threeparttable}
\caption{\label{tab:weights} Noise measurements for the final images. Final weights and tapers used to image each spw. }
\begin{tabular}{cccc}
\noalign{\smallskip} \hline \noalign{\medskip}
Central Frequency &  Noise  &  Taper  & Weight\tnote{a}\\ (MHz) & (mJy beam$^{-1}$) & (k$\lambda$)\\
\noalign{\smallskip} \hline \noalign{\medskip} 
1025 &  0.15 & - & briggs 0 \\
1089 &  0.12 & - & briggs 0.5 \\
1153 &  0.12 & - & briggs 0.5 \\
1217 &  0.13 & - & briggs 0.5 \\
1281 &  0.10 & - & briggs 0.8 \\
1345 &  0.10 & - & natural \\
1409 &  0.14 & - & natural \\
1473 &  0.12 & - & natural \\
1525 &  0.30 & - & natural \\
1589 &  0.20 & 7 & natural \\
1653 &  0.15 & 12 & natural \\
1717 &  0.14 & 12 & natural \\
1781 &  0.12 & 10 & natural \\
1845 &  0.11 & 10 & natural \\
1909 &  0.12 & 10 & natural \\
1973 &  0.09 & 7 & natural \\

\noalign{\smallskip} \hline \noalign{\medskip}
\end{tabular}
\begin{tablenotes}
\item [a] Numbers next to briggs denote the robustness parameter, where -2 maps to uniform weighting and 2 maps to natural weighting
\end{tablenotes}
  \end{threeparttable}
\end{table}

\subsection{Ancillary Data}\label{ancillary}

\subsubsection{VLA HI Data}

To trace the neutral atomic gas of the system we use VLA 21 cm observations originally published in \citet{Mundell2004}. The data were taken in 1995 using the VLA in B configuration, achieving a final restoring beam size of 6.3'', a velocity resolution of 10.3 km/s and an rms per channel of 0.3 mJy/beam. 

\subsubsection{Optical continuum Data}

We use $B$-band observations of the Arp 94 system from the Wide Field Camera (WFC) on the 2.5 m Isaac Newton Telescope (INT) in La Palma, Canary Islands, as presented in \citet{Mundell2004}. The pixel scale for these observations is 0.33'', and the seeing was $\sim$1''.

\subsubsection{H$\alpha$ Data}

We make use of narrowband H$\alpha$ observations of NGC 3227, originally presented in \citet{Mundell2004}, using the 4.2m William Herschel Telescope in La Palma. The observations utilized the TAURUS filter, which has a bandwidth of 15 $\mathring{\text{A}}$. The pixel size in these observations is 0.28".

\subsubsection{ALMA CO(2-1) Data}

We use ALMA CO(2-1) observations of the TDG  published in \citet{Querejeta2021}. The observations combine 12m + 7m + TP array observations, resolving emission down to 0.64'' $\sim$ 45 pc spatial resolution. The 7m array data has a resolution of $\approx$ 5.4'' and a largest recoverable scale of $\approx$ 29'', while the total power array data has a resolution of $\approx$ 28" and is sensitive to emission spanning all scales.

\section{Results}\label{results}

\subsection{Radio Continuum Map}

\begin{figure}
   \centering
   \includegraphics[width=\hsize]{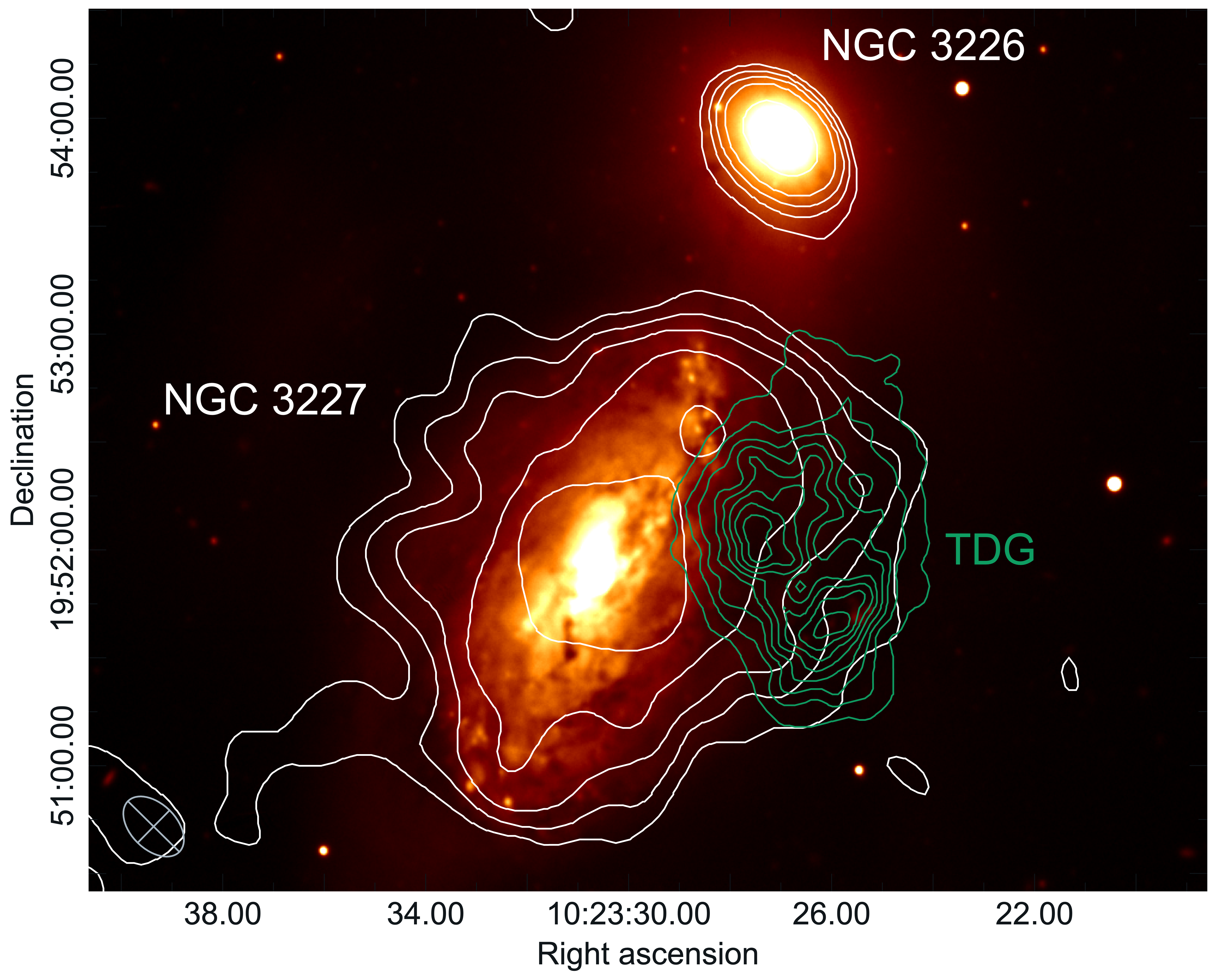}
      \caption{Map of the radio emission of the Arp 94 system in the L-band, combining all spws. The radio contours (white) are superimposed on a $B$-band image. Green contours represent the HI intensity of the TDG observed with VLA \citep{Mundell2004}. Radio contour levels are set at 1.5, 3, 5, 9, and 25 times the rms noise level, with a mean value of 80 $\mu$Jy. The angular resolution is 20'' x 15''.
              }
         \label{FigRadioContours}
   \end{figure}

As described in Section \ref{imaging}, we generated maps of the radio emission between 1-2 GHz for the Arp 94 system. In Figure \ref{FigRadioContours}, we present the EVLA L-band data as contours superimposed on an optical image of the system. The presented radio map integrates the emission from all spws. Additionally, contours representing HI emission from the TDG are also shown.  

Analysis of the L-band data reveals that the central AGNs of NGC 3227 and NGC 3226 are bright sources of radio continuum emission within the system. In addition, we identify extended radio continuum emission emanating from the spiral disk of the parent galaxy of the TDG. Notably, a portion of this emission spatially coincides with the position of TDG J1023+1952. However, given the configuration of the system, where the TDG is situated in front of the spiral disk, it is challenging to precisely determine the extent to which the emission originates from the dwarf galaxy. We also note that in certain regions of the dwarf, the radio continuum emission does not exceed the 3 sigma threshold. However, we notice a pronounced asymmetry in the emission distribution within the system, which seems to be extended towards the TDG's location. This asymmetry suggests the possibility of emission originating from the dwarf galaxy, a hypothesis we discuss further in Section \ref{section:symmetry}.

\subsection{Spectral index map}\label{spectral index}

 \begin{figure*}
   \resizebox{\hsize}{!}
            {\includegraphics[]{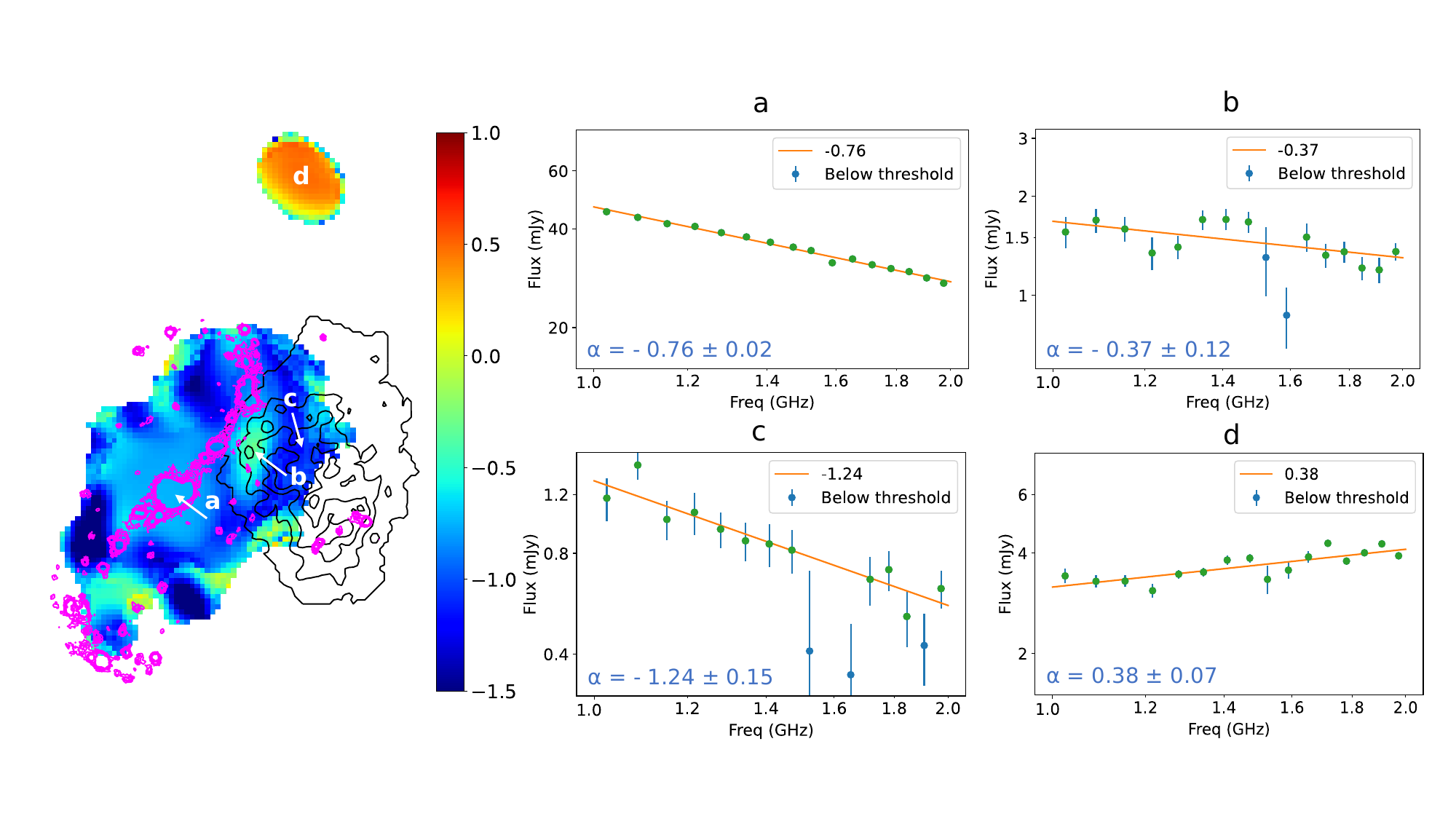}}
      \caption{Spectral Index Map of the Arp 94 system derived from EVLA L-band data.  Pink contours depict H$\alpha$ emission, while black contours represent the TDG's HI emission \citep{Mundell2004}. Specific pixels (a-d) on the spectral index map are indicated, displaying their respective fits and spectral index values. As indicated in the legends inside the fits, data points with flux values below 5 times the rms are colored in blue and are not considered for the fits.
              }
         \label{spindex2column}
    \end{figure*}

To check if the TDG has left a spectral signature, we study the spectral index map of the system using the 16 spws that constitute the EVLA L-band dataset. Despite the limited frequency range of our dataset, we are able to draw interesting conclusions from the spectral index map. We image each spw with a common resolution of $20" \times 15"$, as detailed in Section \ref{imaging}. Throughout this paper, we adopt $S_\nu \propto \nu^\alpha$ as the definition of the spectral index $\alpha$. The resulting spectral index map for the system is presented in Fig. \ref{spindex2column}. It is worth noting that for the least squares fitting procedure, only pixels displaying flux levels exceeding 5 $\times$ rms of each spw image were considered. In addition, a minimum of 5 valid data points from the 16 spws is required to compute the spectral index for each individual pixel. The dip in the spectra of Fig. \ref{spindex2column}, observed in spws centred at 1525 and 1589 MHz, coincides with an L-band region experiencing high RFI and identified as problematic in the VLA observing manual, thus explaining the poorer data quality despite our cleaning efforts.

\subsubsection{NGC 3227}

In the case of the central AGN of NGC 3227, we observe a spectral index of -$0.76 \pm 0.02$ (region "a" in Fig.\ref{spindex2column}). This index is consistent with the presence of non-thermal synchrotron emission. Synchrotron emission generally presents a steep spectrum with spectral index $\alpha \sim -0.8$, and dominates at lower frequencies ($\nu\leq$ 10 GHz, \citealt{Niklas1997}), which corresponds to our L band dataset. At these frequencies, both synchrotron and inverse-Compton losses constitute the primary mechanisms influencing CRes.

It is noteworthy that the spectrum in star-forming galaxies is usually not uniform, but rather exhibits significant radial variation, becoming steeper towards their outer regions as CRes age while propagating away from the star-forming regions where they originate  \citep[e.g.][]{Beck2007,Paladino2009}. 

We can see this steepening away from the star-forming disk at many locations in the spiral galaxy NGC 3227. An example is the region "c" of Fig. \ref{spindex2column}, situated in the outskirts of the galaxy, where the measured spectral index is $\alpha = -1.24 \pm 0.15$. In addition, there are some fluctuations towards the borders of the map which are most likely due to noise effects and will not be further discussed.

\subsubsection{TDG J1023+1952}\label{sec:spindexTDG}
Overall, the spectral index map corresponding to the TDG's location (indicated by the black HI contours in Fig. \ref{spindex2column}) does not provide additional insights into whether the radio continuum emission we are detecting originates from the TDG itself or if we are solely capturing emission from the periphery of the spiral disk of NGC 3227, which lies in the background.

However, it is worth highlighting an intriguing region, denoted as "b" in Fig. \ref{spindex2column}, where the spectrum appears flatter than what would be anticipated for a region not coinciding with intense star formation, as indicated by a lack of H$\alpha$ emission. This anomaly does not seem to be a result of noise or artifacts, as the emission in this region is notably strong and well above the noise threshold. Thus, it is plausible that this anomalous region represents a site where CRes are being accelerated due to the energetic shocks resulting from the collision of the system's tidal streams. A more detailed explanation of this scenario is provided in Section \ref{section: shocks}.

\subsubsection{NGC 3226}

In the case of NGC 3226, we observe a spectral index of $\alpha = 0.38 \pm 0.07$. This is in accordance with what we expect in a LINER (low-ionization nuclear emission-line region) galaxy, as previous studies have demonstrated that some of the most prominent flat-spectrum nuclear radio sources have been found in LINER spectra \citep[e.g.][]{Falcke2000}. 

LINER nuclei are frequently associated with compact radio sources, a phenomenon observed in many E/S0 galaxies. Additionally, flat-spectrum radio sources can be attributed to various factors, such as free-free absorption of non-thermal emission and thermal emission from optically thin ionized gas (see \citealt{Nagar2000} and references therein). The complexity of NGC 3226 within the interacting Arp 94 system was more extensively examined in the work by \citet{Appleton2014}.

\subsubsection{Subtraction of point source}

As part of our analysis, we also examine the spectral index map of the system after removing the central AGN of NGC 3227 to visually differentiate the impact of the AGN from other mechanisms operating in the system. We use standard procedures in CASA to subtract the point source in the visibility data. After the subtraction, the extended emission remains unchanged, indicating that the central emission does not significantly affect the emission originating from the TDG's location. As a result, we have opted to include the central AGN in the images presented throughout the paper. A detailed description of the procedure used to subtract this central point source and the resulting maps are presented in Appendix \ref{Appendix:A}.

\section{Discussion}\label{discussion}

\subsection{Constraints on the radio continuum emission from the TDG}

As previously discussed, it is a challenge to precisely measure the radio continuum emission coming from the TDG, primarily due to an overlap in projection with the spiral galaxy disk. Nevertheless, we employ two methods to estimate this emission, which are described in the following sections. 

\subsubsection{Symmetry of the system} \label{section:symmetry}

\begin{figure} 

    \begin{subfigure}{\columnwidth}
        \centering
        \includegraphics[width=\hsize]{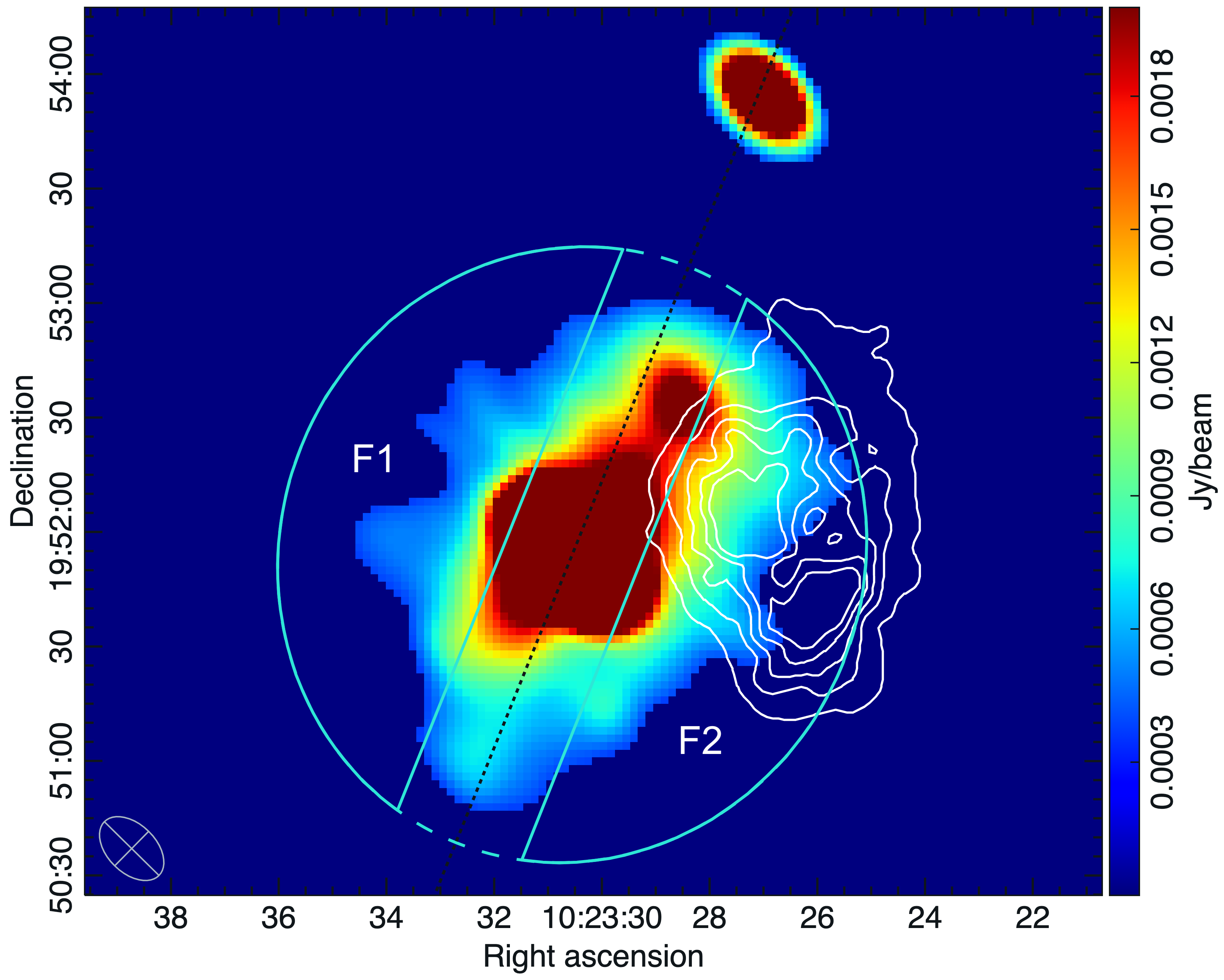}
        \caption{}
        \label{fig:symma}
    \end{subfigure}
    
    \vspace{0.1cm} % Adjust the vertical spacing between subfigures
    
    \begin{subfigure}{\columnwidth}
        \centering
        \includegraphics[width=\hsize]{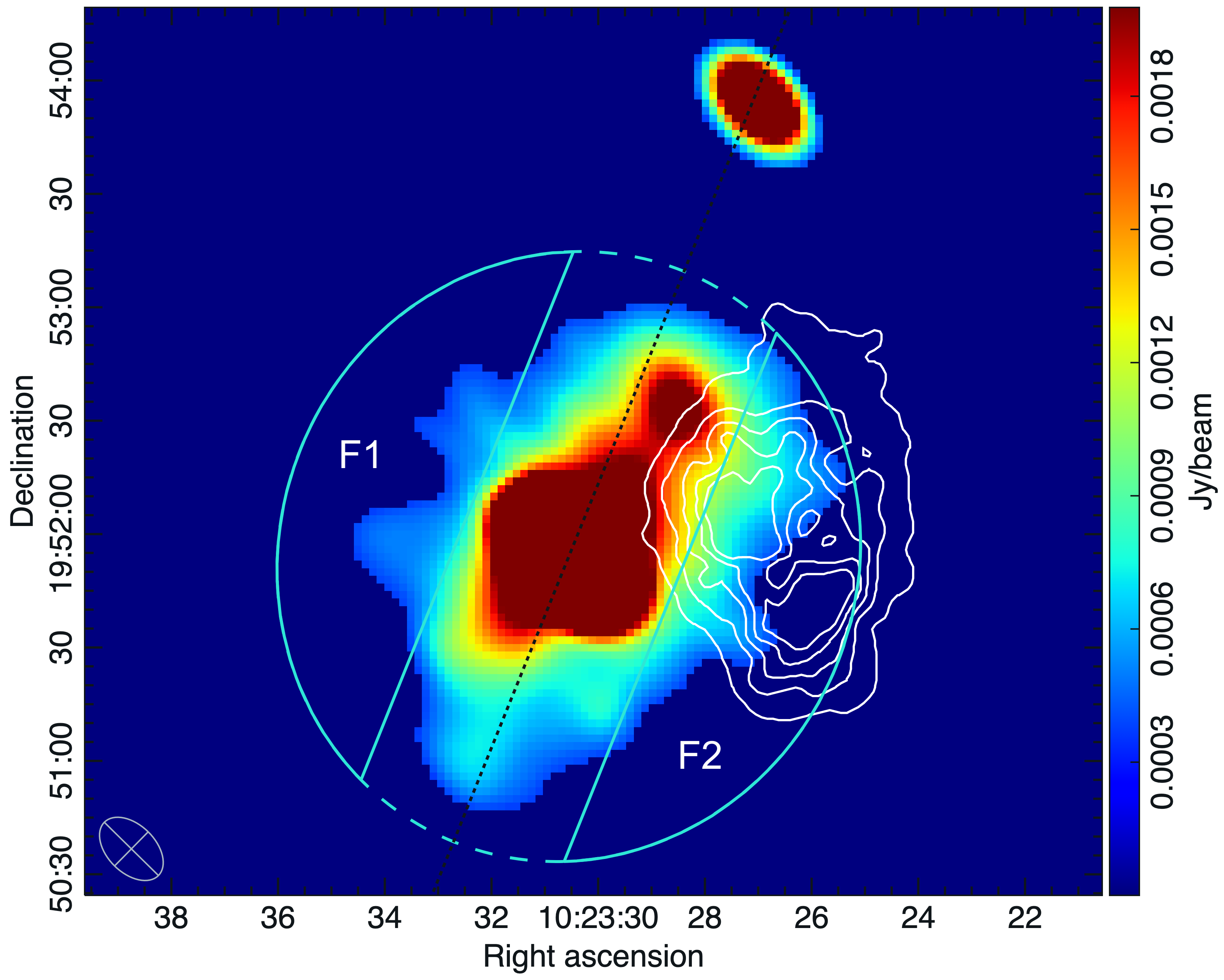}
        \caption{}
        \label{fig:symmb}
    \end{subfigure}
    
    \vspace{0.1cm} % Adjust the vertical spacing between subfigures
    
    \begin{subfigure}{\columnwidth}
        \centering
        \includegraphics[width=\hsize]{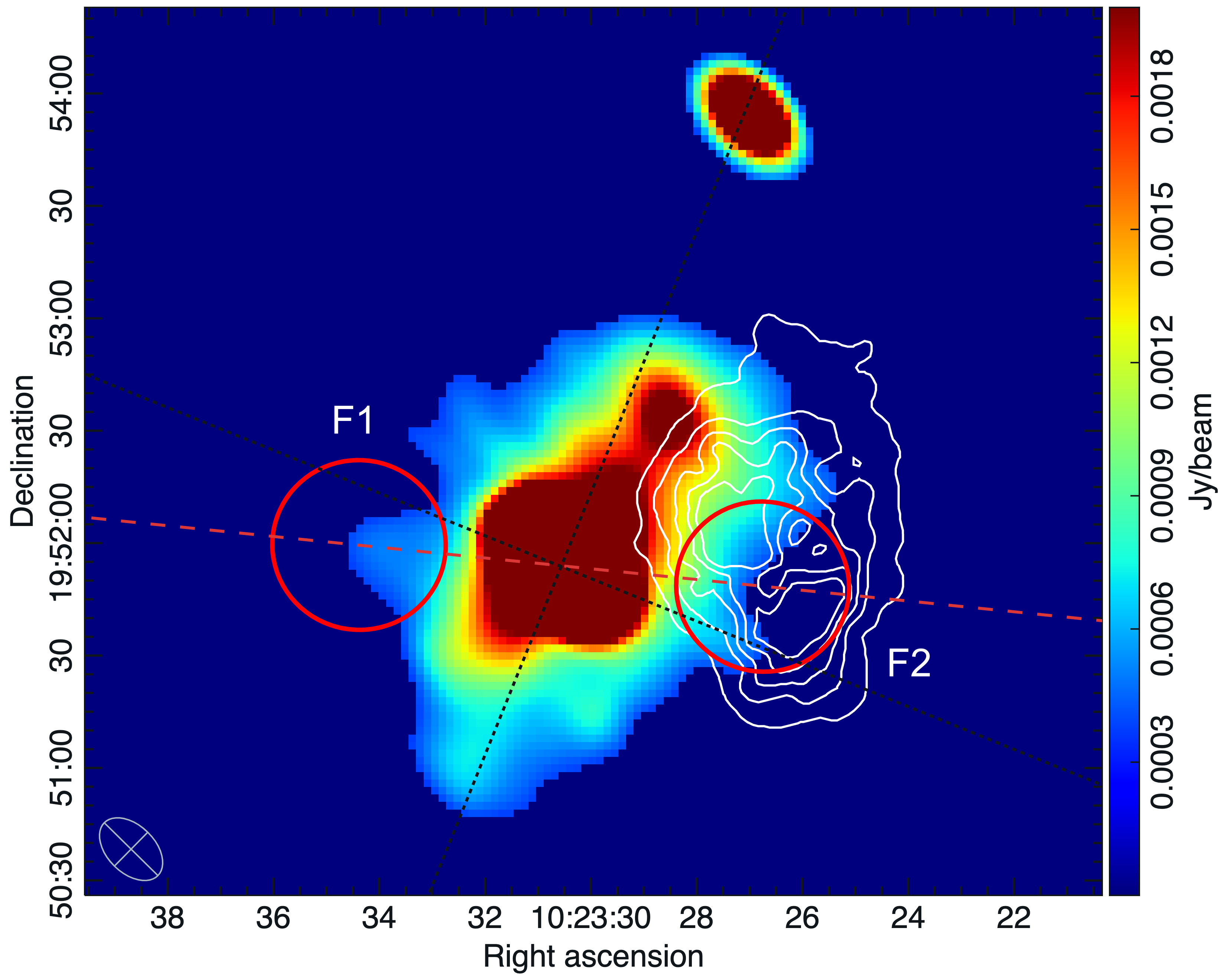}
        \caption{}
        \label{fig:symmc}
    \end{subfigure}
    
    \caption{EVLA L-band radio continuum emission in the Arp 94 system, displaying only emission  $> 5 \sigma$. Panels (a) and (b) illustrate symmetrical regions used for flux comparison, denoted as $F_1$ and $F_2$. The central rectangular mask expands in panel (b). Panel (c) shows two small circular regions symmetrical to the galaxy center.  }
    \label{fig:symmetry}
\end{figure}

Preliminary examination of the radio continuum emission from NGC 3227, depicted in Fig. \ref{FigRadioContours}, reveals an apparent extension towards the TDG's location. To perform a quantitative analysis, we make an estimate based on the assumed symmetry of the emission, where equivalent amounts of radio continuum emission should ideally be measured across both halves of the spiral disk in a perfectly symmetrical system. In this approach, we then tentatively attribute any excess emission on the Western side to the TDG. Naturally, it is not guaranteed that the radio continuum in this object is symmetric, especially since it is interacting. However, the apparent symmetry of the optical disk suggests that it is a reasonable assumption. 

We consider this approach for three different sets of symmetrical regions across NGC 3227, as illustrated in Fig. \ref{fig:symmetry}. In Figs.~\ref{fig:symma} and \ref{fig:symmb} we divide the spiral's disk across its major kinematic axis (PA $=158^\circ \pm 2^\circ$ \citealt{Mundell1995}) and introduce a symmetrical rectangular mask surrounding the emission of the AGN to prevent potential biases arising from NGC 3227's inner regions. In Fig.~\ref{fig:symmb}, we exclude a larger portion of the inner emission. Additionally, we examine the emission within two circular regions in Fig.~\ref{fig:symmc}.
One of these regions is located entirely within the TDG, while the other corresponds to its symmetrical counterpart with respect to the centre of the spiral galaxy.

To eliminate potential biases in the measured flux between regions, we measure the mean flux in blank fields close to both sides of the galaxy. We identify a small difference in the background emission on either side of the system, with a slightly ($\sim$ 0.12 $\mu$Jy/arcsec$^2$) higher value on the side of the TDG.

We quantify the difference in flux between the right, where the TDG is located, ($F_2$) and left ($F_1$) regions, as $\Delta F = F_2 - F_1$ (see Fig. \ref{fig:symmetry}). The results are shown in Table \ref{tab:symmetry}. We estimate the uncertainty of these measurements considering the rms noise measured in emission-free regions of the image, the standard VLA flux calibration uncertainty (approximately $\sim 3\% $ for L-band; \citealt{Perley2017}), and the size of the apertures relative to the beam size.

In all three cases studied, the emission within the regions on the right-hand side of the system, encompassing the TDG, surpasses that within areas in symmetrical regions on the left-hand side, where only the spiral is present. The differences are considerably higher than what could be attributed to the background noise ($\sim$ 0.12 $\mu$Jy/arcsec$^2$). Notably, when examining the normalized emission per unit area, the normalized difference in emission measured within the symmetrical circular regions (Fig. \ref{fig:symmc}) surpasses that observed in the regions of Fig. \ref{fig:symmb}. This could be another indicator that the TDG contributes to the detected radio continuum, since the right circular area is entirely situated within the TDG, unlike the regions in Fig. \ref{fig:symmb} where some emission extends beyond the TDG’s boundaries. 

In addition, since the normalized emission per unit area is roughly consistent among the three sets of regions considered, it can serve as an estimate of the upper limit of the emission coming from the TDG. By multiplying this normalized emission by the area of the TDG, we can derive an upper limit for the emission based on symmetry considerations. In our case, taking the highest normalized emission per unit area, $0.8 \mu\text{Jy/arcsec}^2$, and considering the TDG's area of 64'' $\times$ 43'' \citep{Querejeta2021}, we obtain an upper limit for the total emission of $F_{\text{sym}} < 2.2 \text{ mJy}$.

\begin{table}

\caption{\label{tab:symmetry} Radio continuum flux comparison between symmetrical regions inside and outside the TDG J1023+1952}
\begin{tabular}{ccc}
\noalign{\smallskip} \hline \noalign{\medskip}
Region &  $\Delta F$ (mJy)  &  $\frac{\Delta F}{\text{Area}}$ ($\mu$Jy/arcsec$^2$) \\
\noalign{\smallskip} \hline \noalign{\medskip} 
Large symmetric halves, Fig. \ref{fig:symma} &  $5 \pm 2$ & $0.8 \pm 0.3$ \\
Small symmetric halves, Fig. \ref{fig:symmb} & $3.0 \pm 0.9$ & $0.57 \pm 0.18$ \\
Point symmetric circles, Fig. \ref{fig:symmc} & $ 1.1\pm 0.3$ & $0.67 \pm 0.18$\\
\noalign{\smallskip} \hline \noalign{\medskip}
\end{tabular}

\end{table}

\subsubsection{Upper limits from the rms}\label{upperlim}

We can establish upper limits to the continuum emission by considering the noise levels within the radio map. We define two regions of interest for this analysis: a 30” × 12” rectangle encompassing the star-forming regions and a larger 64” × 43” area encompassing the entirety of the TDG. We display the employed apertures in Appendix \ref{Appendix B}. The computed upper limits are shown in Table \ref{tab:upperlim}, providing valuable constraints about the radio continuum emission associated with a TDG. In this estimate, we make the strong assumption that none of the detected radio emission has its origin in the TDG. Therefore, the upper limits we derive here for the entire TDG using the noise threshold are smaller than the upper limits derived in Section \ref{section:symmetry}, where symmetry arguments were employed.

\begin{table}

\caption{\label{tab:upperlim} Upper limits for the radio continuum emission using a 3$\sigma$ and $5\sigma$ threshold.}
\begin{tabular}{ccc}
\noalign{\smallskip} \hline \noalign{\medskip}
Region &  $F_{3\sigma}$ (mJy)  &  $F_{5\sigma}$ (mJy) \\
\noalign{\smallskip} \hline \noalign{\medskip} 
Star-forming knots &  <0.26 & <0.42 \\
Whole TDG & <0.71 & <1.2 \\
\noalign{\smallskip} \hline \noalign{\medskip}
\end{tabular}
\end{table}

\subsection{Expected radio continuum emission due to star formation}

At frequencies 1 < $\nu$ < 10 GHz the radio continuum emission can be described as the superposition of two main emission mechanisms: thermal free-free emission arising from thermal electrons and non-thermal emission originating from synchrotron radiation. The radio continuum emission can be expressed in terms of these mechanisms as

\begin{equation}
    S_\text{tot}(\nu)= S_{\text{th}}(\nu_0)\left(\frac{\nu}{\nu_0}\right)^{-0.1}+S_{\text{nt}}(\nu_0)\left(\frac{\nu}{\nu_0}\right)^{\alpha_{\text{nt}}}
\end{equation}

\noindent where $\alpha_{\text{nt}}$ is the non-thermal spectral index and -0.1 is the spectral index of optically thin free-free emission. In the following sections, we estimate the different components of the radio continuum emission.

\subsubsection{Thermal Emission}

The relationship between H$\alpha$ and thermal radio continuum emission is well-established, given their common origin in the ionized plasma of HII regions. Several empirical relations in the literature \citep[e.g.][]{Lequeux1980,Murphy2011} have been established to correlate these two quantities. The relation derived by \citet{Lequeux1980} converted to H$\alpha$ takes the form

\begin{equation}
    S_{\text{th, H$\alpha$}} = 1.14\times10^{12}\left(\frac{\nu}{\text{GHz}}\right)^{-0.1}\left(\frac{T_{\rm e}}{10^4\text{K}}\right)^{0.34}\left[\frac{F_{\text{H}\alpha}}{\text{erg s$^{-1}$cm$^{-2}$}}\right] \text{ mJy}
    \label{eq:thermal}
\end{equation}

\noindent where $\nu$ represents the observed frequency in GHz, $T_e$ is the electron temperature (assumed to be $10^4$ K), and $F_{\text{H}\alpha}$ denotes the H$\alpha$ flux.

\citet{Mundell2004} measured the total H$\alpha$ emission coincident with the blue star-forming knots in TDG J1023+1952 using narrowband H$\alpha$ observations of NGC 3227 with the 4.2\,m William Herschel Telescope on La Palma. The total flux measured is $F_{\text{H}\alpha} = 2.55 \times 10^{-14}\text{erg s}^{-1}\text{cm}^{-2}$ uncorrected for extinction. 
We extinction-corrected this flux based on Figure 3b of \citet{Mundell2004}, where the extinction was estimated from the HI column density. In the region of the star-forming knots, the extinction is A(B)= 1.4-2.4 mag. We note that the range of extinctions derived from HI in \citet{Mundell2004} is consistent with the values derived from Spectral Energy Distribution (SED) fitting of individual star-forming knots in \citet{Lisenfeld2008}. We conservatively adopt the upper end of this range and derive A$(\text{H}\alpha) = 1.5$ mag using \citet{Cardelli1989} extinction law. After applying the derived A$(\text{H}\alpha)$, we obtain an extinction-corrected flux of $F_{\text{H}\alpha \text{ corr}} = 1.02 \times 10^{-13}\text{ erg s}^{-1}\text{ cm}^{-2}$.

Using Equation \eqref{eq:thermal}, we estimate the thermal radio continuum component for the star-forming knots within TDG J1023+1952 to be $S_{\text{th, H$\alpha$}} = 0.1$ mJy. Notably, this value is well below the 3$\sigma$ noise limit of our images, thus making the thermal radio continuum undetectable with the current dataset. Therefore, we cannot expect to detect emission exclusively associated with this thermal component, but also need to consider its non-thermal counterpart. 

\subsubsection{Non-thermal Emission}\label{sec:non-thermal emission}
At frequencies below $\sim$ 10 GHz the non-thermal component of the radio emission typically dominates. The modelling of this non-thermal component is more intricate than its thermal counterpart. Models from the literature \citep[e.g.][]{Condon1990,Tabatabaei2017} generally agree that the non-thermal component contributes to around 80\% of the radio continuum emission in the L-band. However, this percentage can vary depending on the specific system under study and the assumed mean $\alpha_{\text{nt}}$. We adopt $\alpha_{\text{nt}}=-1$ based on the study of the radio continuum SED in a sample of nearby galaxies conducted by \citet{Tabatabaei2017}. With this spectral index, we calculate the expected ratio between thermal and non-thermal radio emission adopting the (slightly different) prescriptions of \citet[their eq. 6]{Condon1990} and \citet[their eq. 8]{Tabatabaei2017}. We derive a non-thermal component at $\nu=1.5$ GHz ranging from 0.7 to 0.9 mJy. We comment on the interpretation of this emission as potentially due to star formation from the TDG in Sect. \ref{sec:total radio emission}. 

As a caveat, we note that this estimate of $S_{\rm nt}$ from $S_{\rm th}$ is derived for galaxies with a relatively constant SFR in the past $\sim$ 100 Myr, which is not the case for this TDG, undergoing recent SF. In particular, \citet[][their Section 3.3.2]{Lisenfeld2008} derived with  SED fitting that 6 of the 7 star-forming knots are exceptionally young (< 10 Myr). This suggests that we may be overestimating the synchrotron emission of our sources when applying the general formulas for non-thermal emission \citep{Condon1990,Tabatabaei2017} because the number of supernovae accelerating relativistic electrons in their shocks could be lower than expected from a steady-state situation.

\subsubsection{Total Radio Emission}\label{sec:total radio emission}

By combining the thermal and non-thermal components, we can estimate the total radio continuum emission that we would expect to detect from the TDG. Assuming the thermal component to be 0.1 mJy and the non-thermal component to fall within the range of 0.7 to 0.9 mJy, we anticipate the total radio emission to span $S_\text{1-2GHz}\sim[0.8, 1]$ mJy. This estimate is consistent with other empirical relations found in the literature. For instance, relations derived by \citet{Murphy2011} linking $L_{\text{H}\alpha}$ and $L_{\text{1.4 GHz}}$ suggest a value of $S_\text{1.4 GHz}=0.89$ mJy, which is consistent with our estimated range.

This estimated radio emission (which, in reality, represents an upper limit given the young age of the star clusters as explained in Section~\ref{sec:non-thermal emission}), is close to the estimate derived in Section~\ref{section:symmetry} for symmetric circular apertures, $\sim 1.1$ mJy, covering the southern, star-forming part of J1032+1952. Thus, this tentative emission, or part of it, could indeed originate from the SF activity of the knots, mainly due to synchrotron emission.

However, the radio emission estimated in Section~\ref{section:symmetry} for the whole TDG, of up to 2.2 mJy, cannot be solely attributed to star formation, which appears too faint (only $\lesssim $ 0.9 mJy). If this asymmetric emission stems from the entire TDG (and not from the background spiral), then it should have alternative sources, such as being produced in situ by reacceleration from interstellar shocks, as suggested in  Section \ref{section: shocks}.

\subsection{Interstellar shocks in TDG J1023+1952?}\label{section: shocks}

As described in Section \ref{sec:spindexTDG}, our investigation of the spectral index map of the system revealed an intriguing region with flat spectrum in the edges of the TDG HI cloud. 
This region is unrelated to H$\alpha$ emission, so it does not seem to trace recent star formation, and may indicate a region where CRes have recently undergone reacceleration. 

\citet{Lisenfeld2008} proposed that J1023+1952 is located at the intersection of two tidal streams originating from the Arp 94 system (see Fig. 11 of their paper). They also suggested that the TDG's formation could be a result of this tidal stream interaction, which would explain the high metallicity and molecular gas content of the TDG due to the recycled nature of the tidal stream gas. This scenario, where the TDG's formation is linked to tidal streams, offers a possible explanation for the intriguing flat spectrum observed. If the crossing tidal streams are colliding, the kinetic energy liberated in the hydrodynamic gas interaction could produce shock waves capable of accelerating CRes, resulting in the observed flattening of the spectrum. This process would thus be similar to that described by \citet{Lisenfeld2010} for the bridges formed between colliding pairs of galaxies.

\section{Conclusions}\label{conclusions}

In this work, we have presented new EVLA radio continuum observations in C-configuration and L-band at a resolution of 20'' $\times$ 15'' ($\approx$ 1.2 kpc) for the Arp 94 system, where TDG J1023+1952 is located. Our main findings from the data can be summarized as: 
\begin{enumerate}
    \item We observed an excess of radio continuum emission in the western half of the NGC 3227 disk, coincidental with the location of TDG J1023+1952. Using this asymmetry, we estimated a maximum emission value of 2.2 mJy that might be associated with the TDG. However, this entire emission cannot solely be attributed to star formation in the TDG; it likely has alternative sources such as being generated in situ by the acceleration of CRes in intergalactic shocks. Due to the complex interacting nature of the system, it remains uncertain whether this emission originates from the TDG or the spiral disk.

    \item We derived an upper limit for the emission from the star-forming regions of the TDG of 0.26 mJy with a confidence level of 3 $\sigma$.

    \item Using empirical relations from the literature, we estimated the expected radio continuum emission from the TDG. We estimate from the observed, extinction-corrected H$\alpha$ emission in the star-forming regions, a thermal radio emission of 0.1 mJy and a synchrotron emission of 0.7-0.9 mJy (which might be an upper limit given the young age of the SF knots). This emission could be responsible for part of the observed asymmetry in the radio continuum emission.  
   
    \item We identified a region at the edges of the TDG with a flat spectrum (marked as position "b" in Fig. \ref{spindex2column}). This region is not associated with high H$\alpha$ emission and may suggest gas collisions that accelerate CRes within the tidal tails of TDG J1023+1952.
    
\end{enumerate}
\section*{Acknowledgements}
We thank the anonymous referee for the detailed comments which helped us improve the manuscript.
We wish to thank Prof. Elias Brinks for useful discussions and detailed comments on the manuscript. BMMC acknowledges funding from the Instituto de Astrofísica de Canarias under the IAC 2021 International Scholarships Program. UL acknowledges support by the research project  PID2020-114414GB-I00 financed by MCIN/AEI/10.13039/501100011033, and the Junta de Andalucía (Spain) grant FQM108. MQ acknowledges support from the Spanish grant PID2022-138560NB-I00, funded by MCIN/AEI/10.13039/501100011033/FEDER, EU. The National Radio Astronomy Observatory is a facility of the National Science Foundation operated under cooperative agreement by Associated Universities, Inc. This research made use of the \emph{Cube Analysis and Rendering Tool for Astronomy} (CARTA, \citealt{carta}). This work made use of the following Python libraries: Astropy \citep{Astropy}, Matplotlib \citep{matplotlib}, Numpy \citep{numpy}.

%%%%%%%%%%%%%%%%%%%%%%%%%%%%%%%%%%%%%%%%%%%%%%%%%%
\section*{Data Availability}

Raw data from the EVLA is publicly available in the VLA Archive (\url{https://data.nrao.edu/portal}) under project code 13A-066.
Reduced data will be shared on request
to the corresponding author.



\bibliographystyle{mnras}
\bibliography{paper_1} % if your bibtex file is called example.bib

%\begin{thebibliography}{99}
%\end{thebibliography}

%%%%%%%%%%%%%%%%%%%%%%%%%%%%%%%%%%%%%%%%%%%%%%%%%%

%%%%%%%%%%%%%%%%% APPENDICES %%%%%%%%%%%%%%%%%%%%%

\appendix

\section{Subtraction of central point source from the radio maps}\label{Appendix:A}

We subtracted the central point source in NGC 3227 to assess whether the central emission from the AGN has any impact on the spectral index map of the spiral galaxy. To perform this subtraction, we identified the position and flux of the central point source using our previously generated radio maps, fitting a 2D Gaussian function to a point-like source. The information about the point source was then included in a component list and inserted into the MODEL column of the measuring set. Lastly, we used the CASA \texttt{uvsub} task to subtract the point source model from the calibrated visibilities. 

\begin{figure}
   \centering
   \includegraphics[width=\hsize]{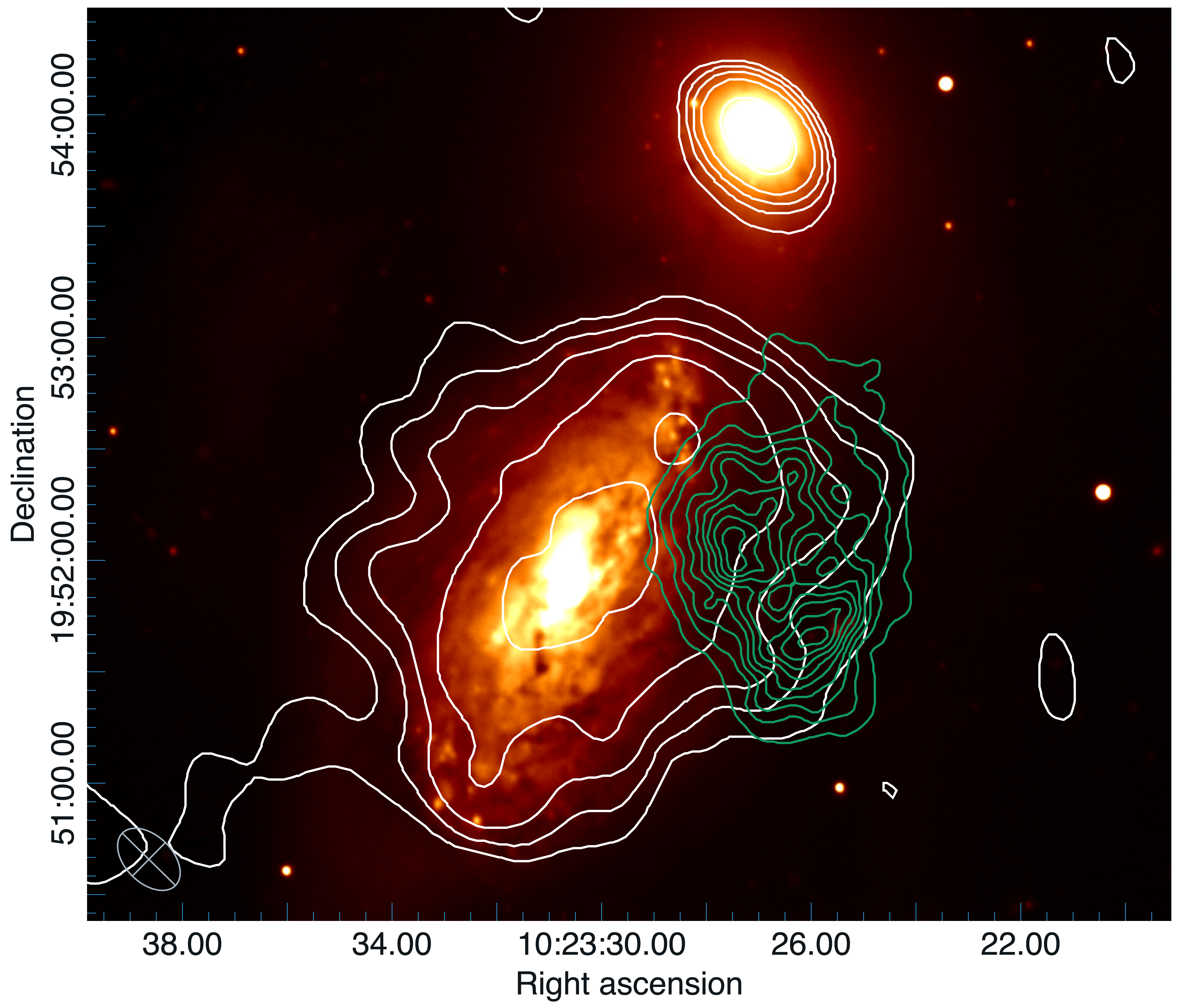}
      \caption{Map of the radio continuum emission of the Arp 94 system in L-band, combining all spws, and subtracting the emission of the central AGN of NGC 3227. The radio contours (white) are superimposed on a B-band image. Green contours represent the HI intensity of the TDG observed with VLA \citep{Mundell2004}. Radio contour levels are set at 1.5, 3, 5, 9, and 25 times the root mean square (rms) noise level, with a mean value of 80 $\mu$Jy. The angular resolution is 20” x 15”.
              }
         \label{FigRadioContours_noagn}
   \end{figure}

In Figure \ref{FigRadioContours_noagn} we present the radio continuum emission in the Arp 94 system, after subtracting the emission generated by the central AGN. This emission is a combination of all spws in the L-band. Notably, the extended emission remains unaltered after the central AGN subtraction, confirming that we can effectively study the emission originating from where the TDG is located even in the presence of the AGN.

\begin{figure}
   \centering
   \includegraphics[width=\hsize]{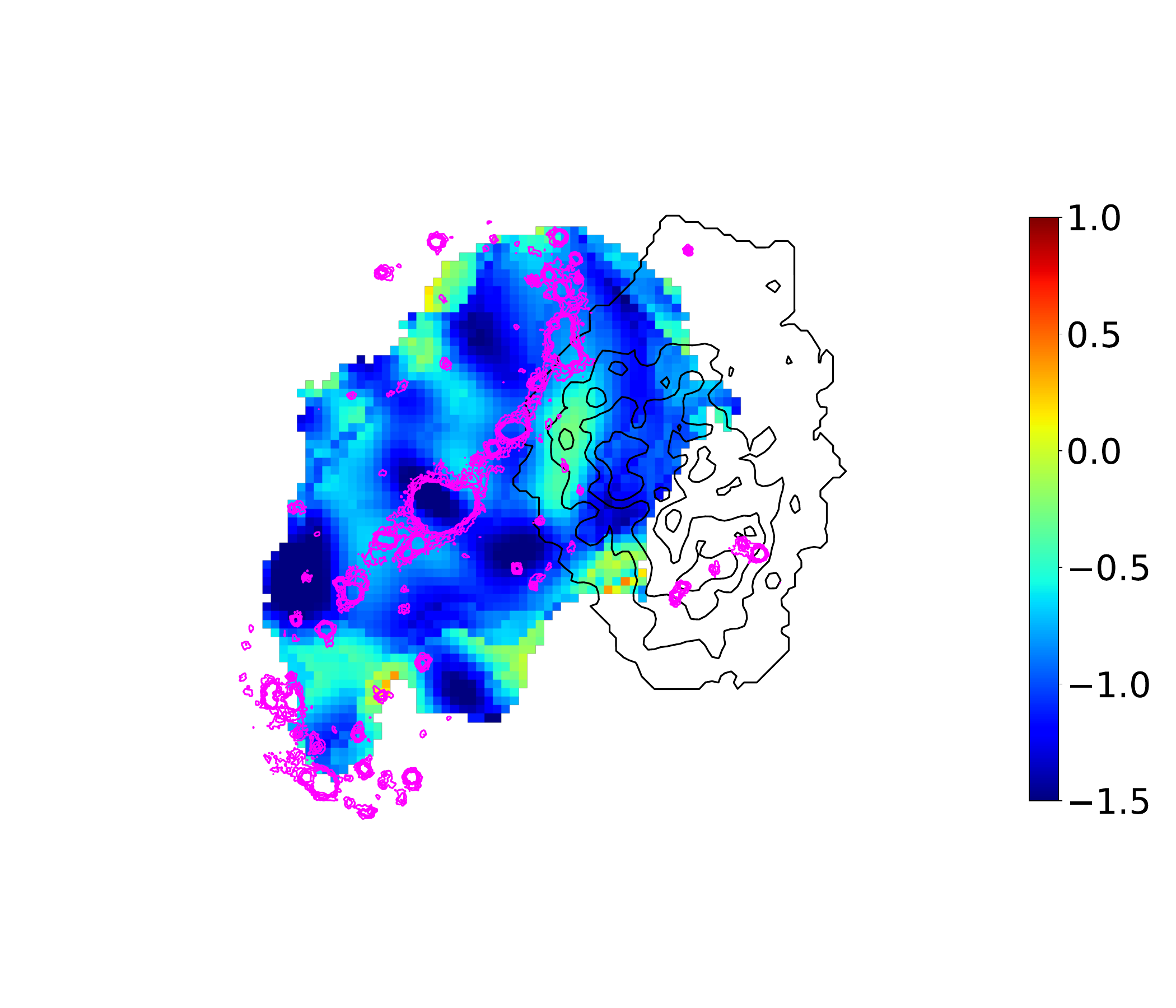}
      \caption{Spectral index map of NGC 3227 after subtracting the central AGN. Pink contours depict H$\alpha$ emission, while black contours represent the TDG’s HI emission \citep{Mundell2004}. 
              }
         \label{Figspindex_noagn}
   \end{figure}

Similarly, in Figure \ref{Figspindex_noagn}, we have examined the spectral index map of NGC 3227 after eliminating the AGN emission. With the exception of the central region, where the spectrum becomes steeper after removing the central AGN, the results are identical to those obtained before subtracting the central point source (see Fig. \ref{spindex2column}). Consequently, it is justifiable to proceed with the analysis of the TDG spectral index without the need to subtract the central AGN.

\section{Selection of regions for rms upper limits}\label{Appendix B}

As described in Section \ref{upperlim}, we used the rms of the radio continuum map to derive upper limits on the radio emission, considering that none of the detected emission originates from the TDG.

\begin{figure}
   \centering
   \includegraphics[width=\hsize]{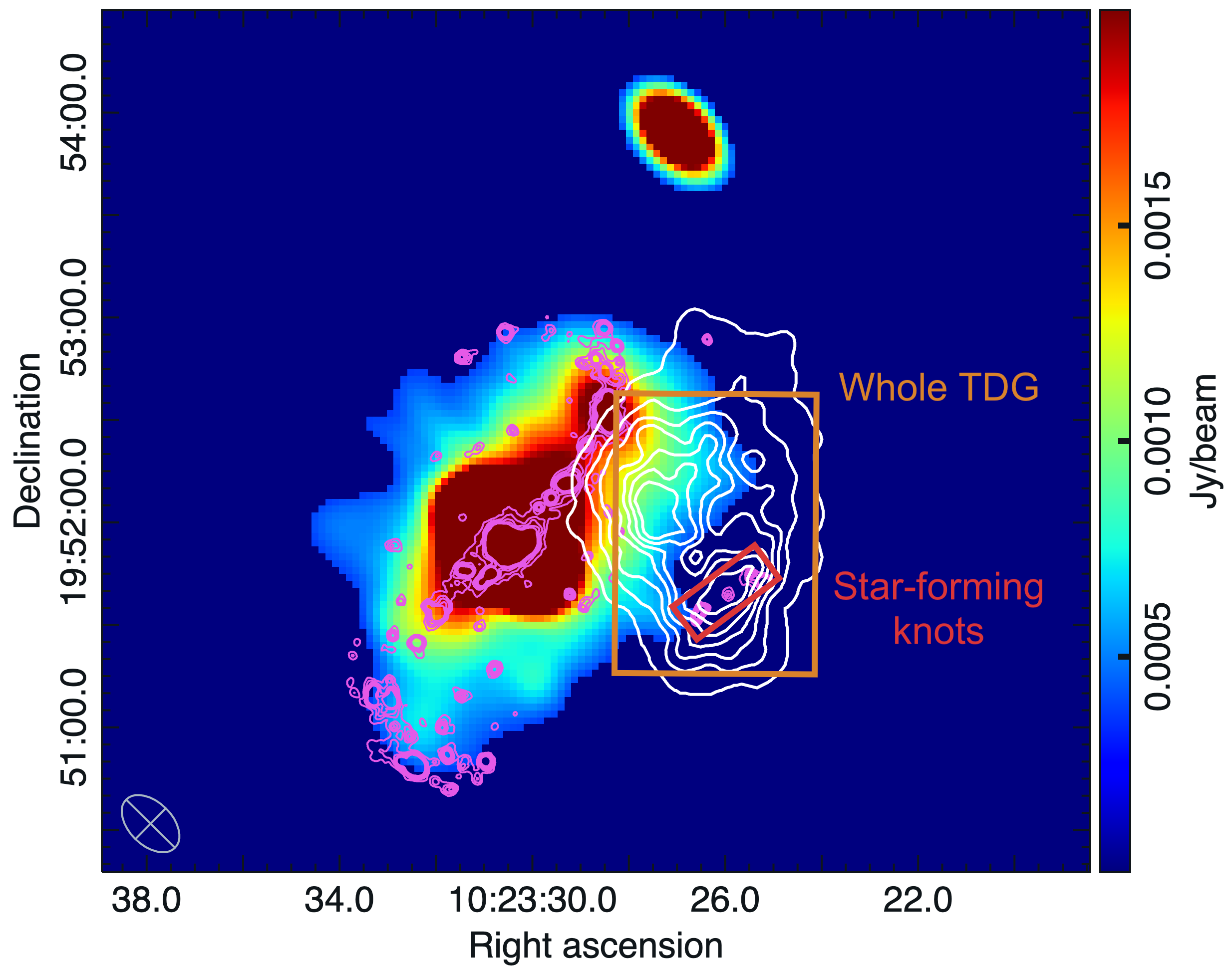}
      \caption{EVLA L-band radio continuum emission in the Arp 94 system, displaying only emission $> 5 \sigma$.  Brown and red rectangles show the apertures used to calculate upper limits as discussed in Section \ref{upperlim}. Pink contours correspond to H$\alpha$ emission, while white contours represent HI emission from the TDG \citep{Mundell2004}.}
              
         \label{Fig:AppendixB_upperlimits}
   \end{figure}

In Fig. \ref{Fig:AppendixB_upperlimits} we illustrate the apertures employed for calculating these upper limits, including a 30” × 12” rectangle encompassing the star-forming knots, as well as a larger area of 64” × 43” covering the whole TDG. The resulting upper limits are shown in Table \ref{tab:upperlim} in the main text.

%%%%%%%%%%%%%%%%%%%%%%%%%%%%%%%%%%%%%%%%%%%%%%%%%%

% Don't change these lines
\bsp	% typesetting comment
\label{lastpage}
\end{document}